\documentclass[preprint]{elsarticle}

\usepackage[utf8]{inputenc}
\usepackage{amsmath}
\usepackage{amsfonts}
\usepackage{amssymb}
\usepackage{amsthm}
\usepackage{bm}
\usepackage{graphicx}
\usepackage{xcolor}
\usepackage[algoruled,linesnumbered]{algorithm2e}
\usepackage{setspace}
\usepackage{geometry}
\usepackage{hyperref}

\hypersetup{%
    colorlinks=true,
    }
    
\newcommand{\edtlang}[1]{#1}

\newcommand{\edt}[1]{#1}

\newcommand{\redt}[1]{#1}

\usepackage{pgfplots}
\pgfplotsset{
    compat=newest,
    every axis plot/.append style={very thick},
    every axis/.append style={label style={font=\large},
                              tick label style={font=\large},
                              legend style={font=\large}}
}

\usepackage{ulem} 

\journal{Signal Processing}

\newtheorem{assumption}{Assumption}
\newtheorem{remark}{Remark}

\newcommand{\ols}[1]{\mskip.5\thinmuskip\overline{\mskip-.5\thinmuskip {#1} \mskip-.5\thinmuskip}\mskip.5\thinmuskip}

\def\RR{\mathbb{R}}
\def\CC{\mathbb{C}}

\newcommand{\vect}[1]{\mathbf{#1}}

\def\x{\vect{x}}
\def\y{\vect{y}}

\def\bSigma{\bm\Sigma}

\def\btheta{\bm\theta}

\newcommand{\matr}[1]{\mathbf{#1}}
\def\ana{\matr{U}} 
\def\C{\matr{C}}
\def\D{\matr{D}}
\def\Mr{\matr{M}}  
\def\Mm{\ols{\Mr}} 
\def\H{\matr{H}}
\def\Id{\matr{I}}  
\def\P{\matr{P}}
\def\syn{\matr{T}} 

\def\W{\matr{W}}

\def\coef{\vect{s}}
\def\COEF{\matr{S}} 
\def\coefh{\hat\coef}
\def\COEFH{\hat\COEF}

\def\sig{\y}
\def\sigh{\hat\sig}

\def\fobs{\x^\textup{obs}}
\def\FOBS{\matr{X}^\textup{obs}}

\def\fmiss{\x^\textup{miss}}
\def\FMISS{\matr{X}^\textup{miss}}
\def\FMISSh{\hat{\matr{X}}^\textup{miss}}
\def\fmissh{\hat\x^\textup{miss}}
\def\fsig{\x}
\def\FSIG{\matr{X}} 
\def\fsigh{\hat\fsig}
\def\FSIGH{\hat\FSIG}

\def\bSigmah{\hat{\bSigma}}

\def\Tdim{L} 
\def\Fdim{F} 
\def\TFdim{{\Fdim \times N}} 
\def\Wdim{M} 

\newcommand{\norm}[1]{\|#1\|}
\newcommand{\abs}[1]{\left\vert#1\right\vert}
\newcommand{\transp}[1]{#1^\mathsf{T}}
\newcommand{\adjoint}[1]{#1^\mathsf{H}}
\newcommand{\inv}[1]{#1^{-1}}
\newcommand{\pinv}[1]{#1^+}
\renewcommand{\det}[1]{\operatorname{det}\left(#1\right)}
\newcommand{\argmin}{\mathop{\operatorname{arg\,min}}}

\newcommand{\diag}{\mathrm{diag}}

\def\EMTF{EM-\textit{tf}}
\def\EMT{EM-\textit{t}}
\def\AM{AM}
\def\AMEM{AM-to-EM-\textit{tf}}

\begin{document}
\sloppy
\begin{frontmatter}

\title{Algorithms for audio inpainting based on probabilistic nonnegative matrix factorization}

\tnotetext[mytitlenote]{%
	The work was supported by the Czech Science Foundation (GAČR) Project No.\,20-29009S, the French ANITI Project No.~ANR-19-PI3A-0004 and the European Research Council (ERC) Project FACTORY No.\,6681839.
	Part of the work was realized during the stay of O. Mokrý at IRIT, co-financed from the Erasmus+ mobility program.
	The authors would like to thank Pavel Rajmic and Pierre-Hugo Vial for their inputs concerning both the development of the methods and the preparation of the manuscript.
}

\author[but]{Ondřej Mokrý\corref{mycorrespondingauthor}}
\cortext[mycorrespondingauthor]{Corresponding author}
\ead{ondrej.mokry@vut.cz}

\author[inria]{Paul Magron}
\ead{paul.magron@inria.fr}

\author[isae]{Thomas Oberlin}
\ead{thomas.oberlin@isae-supaero.fr}

\author[irit]{Cédric Févotte}
\ead{cedric.fevotte@irit.fr}

\address[but]{Brno University of Technology, Faculty of Electrical Engineering and Communications, Department of Telecommunications, Technická 12, 616 00 Brno, Czech Republic}
\address[inria]{Université de Lorraine, CNRS, Inria, LORIA, F-54000 Nancy, France}
\address[isae]{ISAE-SUPAERO, Université de Toulouse, France}
\address[irit]{IRIT, Université de Toulouse, CNRS, Toulouse, France}

\begin{abstract}
Audio inpainting, i.e., the task of restoring missing or occluded audio signal samples, usually relies on sparse representations or autoregressive modeling.
In this paper, we propose to structure the spectrogram with nonnegative matrix factorization (NMF) in a probabilistic framework. First, we treat the missing samples as latent variables, and derive two expectation--maximization algorithms for estimating the parameters of the model, depending on whether we formulate the problem in the time- or time-frequency domain. Then, we treat the missing samples as parameters, and we address this novel problem by deriving an alternating minimization scheme.
We assess the potential of these algorithms for the task of restoring short- to middle-length gaps in music signals. Experiments reveal great convergence properties of the proposed methods, as well as competitive performance when compared to state-of-the-art audio inpainting techniques.
\end{abstract}


\end{frontmatter}


\section{Introduction}

Audio inpainting \cite{Adler2012:Audio.inpainting} is an inverse problem aiming at restoring audio signals degraded by sample loss.
Such a problem typically occurs as a result of packet loss during transmission (packet loss concealment \cite{Lindblom2002:PLC.sinusoidal, Rodbro2006:Packet.Loss.Concealment}) or in digitization of physically degraded media.
Inpainting can also be used to restore signal samples subject to a degradation so heavy that the information about the samples can be considered lost. 
Formally, let $\sig\in\CC^\Tdim$ denote the original time-domain signal prior to the degradation (we consider complex-valued signals for the sake of generality).
The goal of inpainting is to estimate $\sigh\in\CC^\Tdim$, given an incomplete observation of $\sig$.
This problem is ill-posed because of the missing samples, and even the observed samples are prone to some measurement error or noise.
Restoring the signal thus requires some
\edt{assumptions on the structure of} 
the original signal,
in order to guide the estimation towards the most desirable solution.

One of the earliest, yet most successful approaches to audio inpainting is to assume the underlying autoregressive (AR) nature of clean audio signals: Janssen's iterative method \cite{javevr86} is still among state-of-the-art methods to date.
Great performance is also achieved by Etter's extrapolation-based technique \cite{Etter1996:Interpolation_AR}, however, it is limited by the need for a sufficiently long clean context to reconstruct each individual gap.
Recently, the class of sparsity-based methods methods has emerged \cite{Adler2012:Audio.inpainting,MokryRajmic2020:Inpainting.revisited,MokryZaviskaRajmicVesely2019:SPAIN,TaubockRajbamshiBalasz2021:SPAINMOD}.
Generally speaking, these methods solve a regularized inverse problem, where the solution is assumed to have a sparse time-frequency (TF) spectrum, while fitting the observed temporal samples.

A disadvantage of most methods is the local nature of the regularizers.
For example, sparsity-based methods are effective for inpainting of short- to medium-length gaps, typically up to 50\,ms, or for restoring randomly subsampled signals \cite{Lieb2018:Audio.Inpainting,MokryRajmic2020:Approximal.operator}.
Even for gaps of length of tens of milliseconds, there is a need for strong regularization that not only exploits the local TF sparsity but also the global structure of audio signals.
A step towards using global properties of audio signals is the so-called social sparsity \cite{kowalski2012social,SiedenburgKowalskiDoerfler2014:Audio.declip.social.sparsity,GaultierKiticGribonvalBertin:Declipping2021,Zaviska2022:Analysis.social.sparsity.arxiv}, where the significant TF coefficients are expected to occur in particular transient or temporal patterns, or recent approaches based on generative deep neural networks \edt{on medium gaps \cite{Marafioti2019:Context.encoder} and long gaps in the range of seconds} \cite{Marafioti2019:DNN.inpainting, Marafioti2021:GACELA}.

In the present paper, we focus on audio inpainting methods \edt{for medium gaps (in the range of tens of milliseconds)} that leverage
the low-rank structure of audio signals in the TF domain. Among low-rank models, nonnegative matrix factorization~(NMF)
has been intensively used for analysis and decomposition, both in machine learning and signal processing \cite{LeeSeung2000:Algorithms.NMF,WangZhang2013:NMF.comprehensive.review,Huang2012:NMF.short.survey.methods.applications}.
Concerning audio, NMF can be used to decompose the signal's power or magnitude spectrogram
as the product of nonnegative matrices $\W\H$, where $\W$ is a dictionary of spectral patterns, and $\H$ contains the temporal activations of these patterns.
Such a decomposition provides a semantically reasonable generative model for audio signals, which is one of the reasons why NMF is among the classical approaches to source separation \cite{virtanen:NMF,Ozerov2010:Multichannel.NMF.audio.source.separation}.
It also benefits from being an unsupervised and interpretable method while being computationally cheap.
NMF has also been used as a prior for the restoration of degraded signals \cite{BilenOzerovPerez2015:declipping.via.NMF,Bilen2018:NTF_audio_inverse_problems}, thus being successfully applied to audio declipping \cite{BilenOzerovPerez2015:declipping.via.NMF,ZaviskaRajmicOzerovRencker2021:Declipping.Survey}.

The main idea of NMF-based signal reconstruction can be summarized as building an estimation problem where the NMF is used as a prior, which requires to estimate parameters given the incomplete data, usually via the expectation--maximization (EM) algorithm \cite{dempster1977maximum}.
Formally, the structure
of the spectrogram is introduced via the probabilistic Gaussian composite model \cite{fevotte2009nonnegative}, which results in applying NMF with the Itakura--Saito divergence.
However, the estimation of the parameters -- the matrices $\W$ and $\H$ -- is a non-trivial problem which can be approached in different ways.
One possibility is to derive a generalized EM algorithm to estimate the factorization of the power spectrogram of the clean signal in the maximum likelihood (ML) sense.
This has been previously presented by Bilen, Ozerov and Pérez \cite{BilenOzerovPerez2015:declipping.via.NMF,Bilen2018:NTF_audio_inverse_problems} with application to audio declipping, where it reaches state-of-the-art performance.
However, the method is known to be computationally very demanding \cite[Sec.\,V.D]{ZaviskaRajmicOzerovRencker2021:Declipping.Survey}, and it represents only one of several possible approaches to treating the missing samples (specifically, they are treated as latent variables).
Furthermore, it has not been compared to the state-of-the-art methods in the audio inpainting setting.

Drawing on the work of Bilen et al., we provide a novel generalization of their EM-based approach, where the missing samples are formally treated as latent variables.
We formulate the estimation problem in both time and TF domains and therefore derive two algorithms, among which one is novel.
As a novel approach, we also propose to treat the missing samples as parameters. This leads to a new estimation problem, for which we derive an alternating minimization (AM) scheme.
Even though we built upon the application in audio inpainting, the core of the work is the conceptual development of novel approaches to the NMF-based modeling in signal restoration.
We conduct experiments on the task of restoring
\edt{short to middle gaps} 
in music signals \edt{(in particular gaps up to 80\,ms)}. These reveal great convergence properties of the proposed methods, as well as competitive performance when compared to state-of-the-art audio inpainting techniques.

The rest of the paper is organized as follows.
We formulate audio inpainting with NMF as an estimation problem in Section \ref{sec:formulation}.
In Section~\ref{sec:mle.1} we 
review and extend the approach by Bilen et al.\ \cite{BilenOzerovPerez2015:declipping.via.NMF}.
In Section~\ref{sec:mle.2} we consider the missing samples as parameters, for which we derive a novel AM estimation approach.
Section \ref{sec:experiments} is devoted to experiments and evaluation of the proposed methods.
Finally, Section \ref{sec:conclusion} concludes the paper.

\section{Problem formulation}
\label{sec:formulation}

For the whole derivation of the methods, it is convenient to divide the signal into windowed temporal frames $\fsig_n\in\CC^\Wdim,\,n=1,\dots,N$, potentially arranged in a matrix
$\FSIG = [\fsig_1,\dots,\fsig_N]$.
In the case of inpainting, we observe the samples $\fobs_n = \Mr_n\fsig_n$ in each frame $n$ and we aim at obtaining an estimate $\fsigh_n$ of the whole frame given these observed samples.
The matrix $\Mr_n$ is constructed from the identity matrix by omitting the rows corresponding to the missing samples, thus the multiplication with $\Mr_n$ shortens the vector, selecting only the observed samples.
Note that in the present work, the \emph{indices} of the missing samples are assumed to be known, thus the matrices $\Mr_n,\,n=1,\dots,N$ are known.
The whole signal estimate $\sigh\in\CC^\Tdim$ is then obtained by folding all $\fsigh_n$ together by the overlap-add procedure,

To propose a probabilistic formulation of the inpainting problem, it is necessary to postulate a statistical model for the data at hand.
Since we aim at promoting the low-rank structure of the TF coefficients $\COEF = [\coef_1,\dots,\coef_N] = [s_{fn}] \in \CC^{\TFdim}$ of the audio signal, we first define the following synthesis model:
\begin{equation}
	\fsig_n = \syn\coef_n,\quad n = 1,\dots,N,
	\label{eq:reconstruction.operator}
\end{equation}
or, in matrix form, $\FSIG = \syn\COEF$, where $\syn\in\CC^{\Wdim\times\Fdim}$ is a linear reconstruction operator.
The matrix $\syn$ typically represents the inverse discrete Fourier transform (DFT).
If the temporal frames are weighted by a windowing function, $\COEF$ represents the short-time Fourier transform (STFT) coefficients of the whole signal $\sig$.
We will discuss the effect of particular choices of $\syn$ later in \ref{ssec:equivalence}.

The low-rank structure of the TF coefficients can be formalized within the following assumptions.

\begin{assumption}[Gaussian coefficients]
	\label{ass:gaussian}
	The time-frequency coefficients
	are treated as conditionally mutually independent,\footnote{Note that temporal Markov NMF models such as in \cite{Smaragdis2014:Static.dnymic.source.separation.NMF} could readily be considered but we use independence as a working assumption for ease of presentation.} and each coefficient follows a complex circular zero-mean Gaussian distribution:
	\begin{equation}
		s_{fn}\sim\mathcal{N}(0, v_{fn}).
	\end{equation}
\end{assumption}

Due to the assumed independence of the individual TF coefficients, we can rewrite the distribution of the spectrum of the $n$-th frame as:
\begin{equation}
	\coef_n \sim \mathcal{N}(\matr{0}, \D_n),\quad\D_n = \diag \left([v_{fn}]_{f=1,\dots,\Fdim} \right),
	\label{eq:distribution.s}
\end{equation}
where the symbol $\mathcal{N}$ from now on denotes the multivariate complex Gaussian distribution.

\begin{remark}
    \label{rem:dft.independence}
    Note that Assumption \ref{ass:gaussian} is stated for the case of a generic linear transform $\syn$ interconnecting the TF coefficients and temporal samples.
    In the common case where the TF coefficients are computed from the temporal samples by the DFT (i.e., $\syn$ represents the inverse DFT operator), and the temporal signal is real-valued, the independence assumption needs to be relaxed:
    The TF coefficients of real-valued signals are independent only in half of the frequency spectrum, the other half being determined via the Hermitian property of the transform.
    However, for the sake of generality, we will assume no particular structure of $\syn$ throughout the article.
\end{remark}

\begin{assumption}[NMF structure of the variances]
	\label{ass:nmf}
	The variance matrix $\matr{V} = [v_{fn}]$ has the low-rank NMF structure:
	\begin{equation}
		v_{fn} = \sum_{k=1}^{K} w_{fk}h_{kn},
	\end{equation}
	where $K$ is small and all parameters are nonnegative.
	This model amounts to $\matr{V} = \W\H$ with $\W$ and $\H$ being $\Fdim \times K$ and $K \times N$ nonnegative matrices, respectively \cite[Sec.\,2.2]{BilenOzerovPerez2015:declipping.via.NMF}.
\end{assumption}

To estimate the parameters $\W,\H$ of the variance matrix, given the observed samples $\FOBS = \{ \fobs_1,\dots, \fobs_N \}$, we employ maximum likelihood (ML) estimation.
The audio inpainting \textit{per se}, i.e., the computation of the missing samples, is then performed explicitly given the estimated parameters and the reliable samples in a way similar to Wiener filtering.

However, there is not a unique way to formulate the ML problem.
The following sections present two approaches that differ in whether the missing samples are treated as latent variables (Section \ref{sec:mle.1}) or explicitly treated as parameters
(Section \ref{sec:mle.2}).
Note that subsection \ref{ssec:emtf}
reformulates the method proposed by Bilen et al.\ in~\cite{Bilen2018:NTF_audio_inverse_problems}, whereas the rest of Section \ref{sec:mle.1} and the whole Section \ref{sec:mle.2} are new contributions.

\section{ML estimation by treating the missing samples as latent variables}
\label{sec:mle.1}

We first present the ML formulation, where the goal is to estimate the parameters $\W,\H$ of the distribution of the restored signal in the TF domain, by minimizing the negative log-likelihood of the observed samples $\FOBS = \{\fobs_1,\dots,\fobs_N\}$, as presented by Bilen et al.\ in \cite{Bilen2018:NTF_audio_inverse_problems}.
The problem can be formalized as
\begin{equation}
	\hat\W, \hat\H = \argmin_{\W,\H} -\log p(\,\FOBS\mid\W,\H\,).
	\label{eq:mle.1}
\end{equation}

This expression can be broken down for a single frame $n$, where the probability $p$
is given by the distribution of $\coef_n$ from Eq.\,\eqref{eq:distribution.s} and by  the linear observation model as
\begin{equation}
    \fobs_n = \Mr_n\fsig_n = \Mr_n\syn\coef_n\sim\mathcal{N}(\matr{0},\Mr_n\syn \D_n \adjoint{\syn}\transp{\Mr}_n),\quad n = 1,\dots,N,
    \label{eq:distribution.fobs}
\end{equation}
where the dependence on $\W,\H$ is contained within the definition of the matrices $\D_n$.

We resort to the EM algorithm \cite{dempster1977maximum}, in line with Bilen et al.\ \cite{Bilen2018:NTF_audio_inverse_problems}.
However, we broaden their work by considering two different settings of the EM algorithm, depending on the domain of the complete data to be estimated.

\subsection{\EMTF}
\label{ssec:emtf}

First, we briefly describe the EM algorithm for the problem \eqref{eq:mle.1}.
The setting is that the \emph{incomplete data} corresponds to the observed reliable signal, i.e., $\FOBS = \{\fobs_1, \ldots, \fobs_N\}$ in the framed time domain.
The \emph{complete data} corresponds to the TF spectrum $\COEF = [\coef_1, \ldots, \coef_N] \in\CC^\TFdim$ of the original signal (which is reflected by the abbreviation \EMTF).
Finally, the parameters to be estimated are $\btheta = \{ \W, \H \}$, and $\tilde{\btheta}$ is the current value of the parameters.

Using this setting, the EM algorithm aims at minimizing the functional:
\begin{equation}
    Q(\btheta,\tilde{\btheta}) = - \int \log p(\COEF \mid \btheta) p(\COEF\mid\FOBS, \tilde{\btheta}) \,\mathrm{d}\COEF
    \label{eq:em.functional}
\end{equation}
by iterating two steps:
\begin{enumerate}
    \item \textbf{E-step}: compute $Q(\btheta,\tilde{\btheta})$,
    \item \textbf{M-step}: update $\tilde{\btheta}$
    by minimizing $Q(\btheta,\tilde{\btheta})$ with respect to $\btheta$.
\end{enumerate}

\noindent
For the E-step,
we obtain
\begin{equation}
    p(\COEF \mid \FOBS, \tilde{\btheta}) = \prod_{n=1}^N p(\coef_n \mid \fobs_n, \tilde{\btheta}) = \prod_{n=1}^N \mathcal{N}(\coef_n\mid\coefh_n, \bSigmah_n),
\end{equation}
where
\begin{subequations}\label{eq:e-step.1}
	\begin{align}
		\coefh_n &= \D_n \adjoint{\syn}\transp{\Mr}_n
		\inv{\left(\Mr_n\syn \D_n \adjoint{\syn}\transp{\Mr}_n\right)}
		\fobs_n,
		\label{eq:e-step.1:mean}
		\\
		\bSigmah_n  &= 
		\D_n - \D_n \adjoint{\syn}\transp{\Mr}_n
		\inv{\left(\Mr_n\syn \D_n \adjoint{\syn}\transp{\Mr}_n\right)}
		\Mr_n\syn \D_n,
		\label{eq:e-step.1:covariance}
	\end{align}
\end{subequations}
and the matrices $\D_n$ are computed using the current value of the parameters $\tilde{\btheta}$.
The formulas \eqref{eq:e-step.1} can be derived from the assumed distribution of $\coef_n$ in \eqref{eq:distribution.s} and from the linear observation model $\fobs_n = (\Mr_n\syn)\coef_n$ (see e.g. \cite[Theorem 10.3]{Kay1993:Fundamentals.Statistical.Processing}). Note that these formulas can be seen as Wiener filtering given estimated posterior covariance~\cite[Chapter 12]{Kay1993:Fundamentals.Statistical.Processing} to recover the missing samples.

Now, from \eqref{eq:distribution.s} and from the independence assumption, it holds that
\begin{equation}
    p(\COEF \mid \btheta) = \prod_{n=1}^N \mathcal{N}(\coef_n \mid \matr{0}, \D_n),\quad\D_n = \diag \left([v_{fn}]_{f=1,\dots,\Fdim} \right).
\end{equation}

The M-step, i.e.,\ the minimization of \eqref{eq:em.functional} with respect to the parameters $\btheta = \{\W,\H\}$ is equivalent to the minimization of
the Itakura--Saito divergence $D_\mathrm{IS}(\P\mid\W\H)$ \cite{BilenOzerovPerez2015:declipping.via.NMF, fevotte2009nonnegative}, where the divergence is defined for matrices $\matr{A} = [a_{ij}],\matr{B} = [b_{ij}]$ as
\begin{equation}
    D_\mathrm{IS}(\matr{A}\mid\matr{B}) = \sum_{i,j} d_\mathrm{IS}(a_{ij} \mid b_{ij}) = \sum_{i,j} \left( \frac{a_{ij}}{b_{ij}} - \log\frac{a_{ij}}{b_{ij}} - 1 \right),
    \label{eq:is}
\end{equation}
and the matrix $\P = [p_{fn}]$ is the posterior power spectrum given by:
\begin{equation}\label{eq:posterior.power}
	p_{fn} = \mathbb{E}\left( \abs{s_{fn}}^2 \mid \fobs_n, \W, \H \right) = \abs{(\coefh_n)_f}^2 + (\bSigmah_n)_{ff}.
\end{equation}
The M-step can thus be performed
by applying (and iterating) the following multiplicative rules \cite[Alg.\,1]{fevotte2009nonnegative}:
\begin{subequations}\label{eq:wh}
	\begin{align}
		\W &\leftarrow \W \odot \frac{%
			\left((\W\H)^{\odot[-2]} \odot \P\right) \transp{\H}
		}{%
			(\W\H)^{\odot[-1]} \transp{\H}
		},
		\label{eq:wh:w}
		\\
		\H &\leftarrow \H \odot \frac{%
			\transp{\W}\left((\W\H)^{\odot[-2]} \odot \P\right)
		}{%
			\transp{\W}(\W\H)^{\odot[-1]}
		},
		\label{eq:wh:h}
	\end{align}
\end{subequations}
where $\frac{\matr{A}}{\matr{B}}$ denotes the matrix $\matr{A}\odot\matr{B}^{\odot[-1]}$ and the symbol $\odot$ is used to denote entry-wise multiplication or power.
Note that in the algorithm, the $\H$ update \eqref{eq:wh:h} uses the already updated value of $\W$ from \eqref{eq:wh:w}.
In practice, the two updates are followed by a normalization step:
The columns of $\W$ are scaled to have unit norm and the rows of $\H$ are inversely scaled by the same factor, such that the product $\W\H$ does not change.

The whole \EMTF{} algorithm is summarized in Alg.\,\ref{alg:emtf}.
Its output is the estimate of the complete data, i.e., the full TF spectrum $\COEFH\in\CC^\TFdim$, together with the estimate of the parameters $\W$ and $\H$.
Then, the framed time-domain signal $\FSIGH$ is synthesized from $\COEFH$ using the operator $\syn$. Finally, the whole signal estimate $\sigh$ is obtained from $\FSIGH$ using the overlap-add procedure.

\begin{algorithm}[t]
    \label{alg:emtf}
	\small
	\DontPrintSemicolon
	\KwIn{reliable samples $\{\fobs_n\}_{n=1,\dots,N}$, respective selection matrices $\{\Mr_n\}_{n=1,\dots,N}$, linear transform $\syn\in\CC^{\Wdim\times\Fdim}$}
	\vspace{0.5em}
	initialize $\W\in\RR^{\Fdim\times K}$, $\H\in\RR^{K\times N}$ non-negative\\
	\Repeat{convergence criteria met}
	{
		\setstretch{1.5}
		\tcp{E-step:}
		\For{$n = 1,\dots,N$}
		{
			$\D_n \leftarrow \diag \left([v_{fn}]_{f=1,\dots,\Fdim} \right)$ with
			$[v_{fn}]_{f=1,\dots,\Fdim}$ being the $n$-th column of the matrix $\matr{V} = \W\H$
			\\
			$\coefh_n \leftarrow \D_n \adjoint{\syn}\transp{\Mr}_n
			\inv{\left(\Mr_n\syn \D_n \adjoint{\syn}\transp{\Mr}_n\right)}
			\fobs_n$ \label{alg:coef}
			\\
			$\bSigmah_n
			\hspace{-.1em}\leftarrow\hspace{-.1em}
			\D_n
			\hspace{-.1em}-\hspace{-.1em}
			\D_n \adjoint{\syn}\transp{\Mr}_n
			\hspace{-.2em}\inv{\left(\Mr_n\syn \D_n \adjoint{\syn}\transp{\Mr}_n\right)}\hspace{-.2em}
			\Mr_n\syn \D_n\hspace{-10em}$ \label{alg:sigma}
			\\
			$p_{fn} \leftarrow \abs{(\coefh_n)_f}^2 + (\bSigmah_n)_{ff},\; f = 1,\dots,\Fdim$

		}
		\tcp{M-step:}
		\Repeat{satisfied with the factorization}
		{
			$\W \leftarrow \W \odot \cfrac{%
				\big((\W\H)^{\odot[-2]} \odot \P\big) \transp{\H}
			}{%
				(\W\H)^{\odot[-1]} \transp{\H}
			}$
			with $\P = [p_{fn}]\hspace{-10em}$
			\\[0.5em]
			$\H \leftarrow \H \odot \cfrac{%
				\transp{\W}\big((\W\H)^{\odot[-2]} \odot \P\big)
			}{%
				\transp{\W}(\W\H)^{\odot[-1]}
			}$
			with $\P = [p_{fn}]\hspace{-10em}$
			\\[0.5em]
			normalize columns of $\W$,
			scale rows of $\H$\\[0.5em]
		}
	}
	\vspace{0.5em}
	\KwOut{$\COEFH =  \left[ \coefh_1,\dots,\coefh_N \right], \hat\W = \W, \hat\H = \H$}
	\caption{Audio inpainting via \EMTF.}
\end{algorithm}

\subsection{\EMT}
\label{ssec:emt}

As an alternative to \EMTF, we build an EM algorithm that uses a complete dataset in the time domain (hence the abbreviation \EMT).
As a result, we directly get the posterior distribution of the missing data as a side-product of the algorithm.

This novel algorithm addresses the original problem \eqref{eq:mle.1} by minimizing the functional:
\begin{equation}
    Q(\btheta,\tilde{\btheta}) = - \int \log p(\FSIG \mid \btheta) p(\FMISS\mid\FOBS, \tilde{\btheta}) \,\mathrm{d}\FMISS,
\end{equation}
where $\FMISS = \{\fmiss_1,\dots,\fmiss_N \}$ represents the missing samples.
These are identified as ${\fmiss_n = \Mm_n\fsig_n}$,
where $\Mm_n$ is the complementary selection matrix to $\Mr_n$, i.e., $\Mm_n$ selects the unrealiable (missing) samples of frame $n$.

It is clear that in some cases, this approach will be equivalent to \EMTF, such as in the case of one-to-one correspondence between the temporal frame samples and their frequency coefficients.
However, this part discusses a general situation with no other assumptions on the transforms involved, and particular cases will be examined later in Section \ref{ssec:equivalence}.

To derive the steps of the EM algorithm, observe that the relation $\fsig_n = \syn\coef_n$, together with Assumption \ref{ass:gaussian} (\nameref{ass:gaussian}), directly leads to
\begin{equation}\label{eq:distribution.x}
    p(\FSIG\mid\btheta) = \prod_{n=1}^N \mathcal{N}\left(\fsig_n \mid \matr{0}, \syn\D_n\adjoint{\syn} \right).
\end{equation}
\edt{%
Using the selection matrices $\Mr_n$ and $\Mm_n$, we derive the distribution
\begin{equation}
	\begin{bmatrix}
		\Mr_n\fsig_n \\ \Mm_n\fsig_n
	\end{bmatrix}
	=
	\begin{bmatrix}
		\fobs_n \\ \fmiss_n
	\end{bmatrix}
	\sim
	\mathcal{N}\left(\matr{0},
	\begin{bmatrix}
		\Mr_n\syn\D_n\adjoint{\syn}\transp{\Mr}_n & \Mr_n\syn\D_n\adjoint{\syn}\transp{\Mm}_n \\[1ex]
		\Mm_n\syn\D_n\adjoint{\syn}\transp{\Mr}_n & \Mm_n\syn\D_n\adjoint{\syn}\transp{\Mm}_n
	\end{bmatrix}
	=
	\begin{bmatrix}
		\C_{11} & \C_{12} \\[1ex]
		\C_{21} & \C_{22}
	\end{bmatrix}
	\right),
\end{equation}
from where it is straightforward to derive that $\fmiss_n$, given $\fobs_n$ and the model parameters, follows the distribution
$\mathcal{N}\left(\C_{21}\inv{\C_{11}}\fobs_n,\,\C_{22} - \C_{21}\inv{\C_{11}}\C_{12} \right)$.
Using the results from Eq.\,\eqref{eq:e-step.1}, we can simply express
$\C_{21}\inv{\C_{11}}\fobs_n = \Mm_n\syn\coefh_n$
and
$\C_{22} - \C_{21}\inv{\C_{11}}\C_{12} = \Mm_n\syn\bSigmah_n\adjoint{\syn}\transp{\Mm}_n$. This yields:
}
\begin{equation}\label{eq:em2.from.em1}
    p(\FMISS \mid \FOBS, \tilde{\btheta}) = \prod_{n=1}^N p(\fmiss_n \mid \fobs_n, \tilde{\btheta}) = \prod_{n=1}^N
    \mathcal{N}(\fmiss_n\mid \Mm_n\syn\coefh_n,\, \Mm_n\syn\bSigmah_n\adjoint{\syn}\transp{\Mm}_n).
\end{equation}

The crucial part of the algorithm is the 
ML estimation of the parameters $\W,\H$ in the M-step, given the posterior distribution of the missing temporal samples.
Even though the closed form of $Q(\btheta,\tilde{\btheta})$ is available due to the expressions in Eq.\,\eqref{eq:distribution.x} and \eqref{eq:em2.from.em1}, it is expensive to compute and optimize directly.
Thus, we proceed to re-estimate the spectrum corresponding to the signal estimated by Eq.\,\eqref{eq:em2.from.em1} and update $\W$ and $\H$ as the factorization of this spectrum.

To do this, we introduce an analysis operator, represented by the matrix $\ana\in\CC^{\Fdim\times\Wdim}$, associated to the synthesis operator $\syn$.
So far, the only assumption about the analysis operator is linearity -- more details are given below.
Using $\ana$, it is straightforward to derive the posterior distribution of the TF coefficients $\COEF^{\textup{alt}} = \ana\FSIG$ associated to the posterior time-domain samples:
\begin{equation}
    p(\COEF^{\textup{alt}} \mid \FOBS, \tilde{\btheta}) = \prod_{n=1}^N p(\coef_n^{\textup{alt}} \mid \fobs_n, \tilde{\btheta}) = \prod_{n=1}^N
    \mathcal{N}(\coef_n^{\textup{alt}} \mid \coefh_n^{\textup{alt}},\, \bSigmah_n^{\textup{alt}})
\end{equation}
with
\begin{equation}\label{eq:e-step.2}
		\coefh_n^{\textup{alt}}	= \ana\syn\coefh_n,
		\quad
		\bSigmah_n^{\textup{alt}} = \ana\syn\bSigmah_n\adjoint{\syn}\adjoint{\ana}.
\end{equation}

Finally, the M-step is equivalent to the M-step defined by the updates in Eq.\,\eqref{eq:wh}, with the alternative posterior power spectrum (computed from $\coefh_n^{\textup{alt}}$ and $\bSigmah_n^{\textup{alt}}$).
The whole algorithm is summarized in Alg.\,\ref{alg:emt}.

\begin{algorithm}[t]
    \label{alg:emt}
	\small
	\DontPrintSemicolon
	\KwIn{reliable samples $\{\fobs_n\}_{n=1,\dots,N}$, respective selection matrices $\{\Mr_n\}_{n=1,\dots,N}$, linear transforms $\syn\in\CC^{\Wdim\times\Fdim}, \ana\in\CC^{\Fdim\times\Wdim}$}
	\vspace{0.5em}
	initialize $\W\in\RR^{\Fdim\times K}$, $\H\in\RR^{K\times N}$ non-negative\\
	\Repeat{convergence criteria met}
	{
		\setstretch{1.5}
		\tcp{E-step:}
		\For{$n = 1,\dots,N$}
		{
			$\D_n \leftarrow \diag \left([v_{fn}]_{f=1,\dots,\Fdim} \right)$ with
			$[v_{fn}]_{f=1,\dots,\Fdim}$ being the $n$-th column of the matrix $\matr{V} = \W\H$
			\\
			$\coefh_n^{\textup{alt}} \leftarrow \ana\syn\D_n \adjoint{\syn}\transp{\Mr}_n
			\inv{\left(\Mr_n\syn \D_n \adjoint{\syn}\transp{\Mr}_n\right)}
			\fobs_n$ \label{alg:coef}
			\\
			$\bSigmah_n^{\textup{alt}}
			\hspace{-.1em}\leftarrow\hspace{-.1em}
			\ana\syn\left(
			\D_n
			\hspace{-.1em}-\hspace{-.1em}
			\D_n \adjoint{\syn}\transp{\Mr}_n
			\hspace{-.2em}\inv{\left(\Mr_n\syn \D_n \adjoint{\syn}\transp{\Mr}_n\right)}\hspace{-.2em}
			\Mr_n\syn \D_n
			\right)\adjoint{\syn}\adjoint{\ana}$
			\\
			$p_{fn} \leftarrow \abs{(\coefh_n^{\textup{alt}})_f}^2 + (\bSigmah_n^{\textup{alt}})_{ff},\; f = 1,\dots,\Fdim$

		}
		\tcp{M-step:}
		\Repeat{satisfied with the factorization}
		{
			$\W \leftarrow \W \odot \cfrac{%
				\big((\W\H)^{\odot[-2]} \odot \P\big) \transp{\H}
			}{%
				(\W\H)^{\odot[-1]} \transp{\H}
			}$
			with $\P = [p_{fn}]\hspace{-10em}$
			\\[0.5em]
			$\H \leftarrow \H \odot \cfrac{%
				\transp{\W}\big((\W\H)^{\odot[-2]} \odot \P\big)
			}{%
				\transp{\W}(\W\H)^{\odot[-1]}
			}$
			with $\P = [p_{fn}]\hspace{-10em}$
			\\[0.5em]
			normalize columns of $\W$,
			scale rows of $\H$\\[0.5em]
		}
	}
	\vspace{0.5em}
	\KwOut{$\COEFH^{\textup{alt}} =  \left[ \coefh_1^{\textup{alt}},\dots,\coefh_N^{\textup{alt}} \right], \hat\W = \W, \hat\H = \H$}
	\caption{Audio inpainting via \EMT.}
\end{algorithm}

\subsection{On the relations between \EMTF{} and \EMT}
\label{ssec:equivalence}

As previously mentioned, there are some natural choices of the pair $\{ \syn,\ana \}$ which result in the equivalence of the algorithms \EMTF{} and \EMT.
Several alternatives are discussed in what follows.
\begin{enumerate}
	\item
	\label{itm:T.is.unitary}
	$\syn$ is invertible, $\ana = \inv{\syn}$.\\
	It follows from \eqref{eq:e-step.2} that $\coefh_n = \coefh_n^\textup{alt}$ and $\bSigmah_n = \bSigmah_n^\textup{alt}$, therefore \EMTF{} and \EMT{} are identical algorithms.
	One noticeable special case is when $\syn$ is unitary, i.e.,\ $\ana = \inv{\syn} = \adjoint{\syn}$.
	A popular example is the case of a properly scaled DFT realized formally by multiplication with the unitary matrix $\ana\in\CC^{\Wdim\times\Wdim}$.
	\item
	\label{itm:T.is.synthesis}
	$\syn$ is the synthesis operator of a tight frame \cite[Ch.\,1]{christensen2008}, $\Fdim > \Wdim$, $\ana = \adjoint{\syn}$, $\syn\ana = \Id$, where $\Id$ denotes the identity matrix of appropriate size.\\
	The two algorithms are no longer equivalent, since $\ana\syn$ represents the projection operator on the range space of $\ana$, which is in general different from identity.
	For instance, this is the case of a redundant DFT, e.g., with $\Fdim = 2\Wdim$ (twice more frequency bins than the frame length).%
	\footnote{%
		In practice, this can be implemented by zero-padding the signal to twice its length and then computing the DFT.
		The backward transform is the inverse DFT, followed by cropping the result to the original length.
	}
	\item $\syn$ is the analysis operator of a tight frame, $\Fdim < \Wdim$, $\ana = \adjoint{\syn}$, i.e., $\ana\syn = \Id$.\\
	In this case, we do not have enough frequency coefficients to reconstruct \emph{any} signal in the framed time domain.
	This means that the time-domain solution (in each frame) is restricted to the range space of $\syn$.
	However, the estimation \EMTF{} and \EMT{} are once again equivalent in this case.
	\item $\syn$ is arbitrary, $\ana = \pinv{\syn}$.\\
	In this case, the matrix $\ana\syn = \pinv{\syn}\syn$ used in \eqref{eq:e-step.2} represents the orthogonal projection onto the range space of $\adjoint{\syn}$ (which equals the orthogonal complement of the kernel of $\syn$).
	This is in general different from identity, unless the range space of $\adjoint{\syn}$ is the whole coefficient space~$\CC^\Fdim$.
	\item $\ana$ is arbitrary, $\syn = \pinv{\ana}$.\\
	Similarly to the previous option, $\ana\syn = \ana\pinv{\ana}$ represents the orthogonal projection onto the range space of $\ana$ (which equals the orthogonal complement of the kernel of $\adjoint{\ana}$).
	This is in general different from identity, unless the range space of $\ana$ is the whole coefficient space $\CC^\Fdim$.
\end{enumerate}

Even though the list is far from being exhaustive, it illustrates that there are commonly used settings (e.g., the redundant DFT used in some sparsity-based reconstruction algorithms \cite{ZaviskaRajmicMokryPrusa2019:SSPADE_ICASSP,MokryZaviskaRajmicVesely2019:SPAIN,TaubockRajbamshiBalasz2021:SPAINMOD}) where \EMTF{} and \EMT{} are not just conceptually but also practically different.
This will be further detailed in the experiments in Section \ref{sec:experiments}.

\section{ML estimation by treating the missing samples as parameters: the \AM{} algorithm}
\label{sec:mle.2}

As a novel approach, we propose to treat the missing samples as parameters and include them explicitly into the estimation problem, which results in:
\begin{equation}
	\hat\W,\hat\H,\FMISSh = \argmin_{\W,\H,\FMISS} -\log p\big( \FOBS, \FMISS \mid \W, \H \big).
	\label{eq:mle.2}
\end{equation}
Note that the objective of this novel estimator is a function different from the likelihood in \eqref{eq:mle.1}, since the parameter space is extended by the inclusion of the missing samples in the problem.
We propose to approach it via AM. This approach consists of two steps -- minimization of \eqref{eq:mle.2} with respect to the missing samples (signal update) and minimization with respect to the NMF parameters (model update).

\subsection{Signal update}

Performing the signal update means minimizing the objective in \eqref{eq:mle.2} with respect to $\fmiss_n$ while the current estimates of $\W$, $\H$ are fixed.
This is equivalent to finding the mode of the conditional distribution of $\fmiss_n$ given $\fobs_n,\W,\H$, which, due to the Gaussian assumption, equals its expectation:
\begin{equation}\label{eq:am.signal.update}
	\fmissh_n
	= \mathbb{E}\left( \fmiss_n \mid \fobs_n, \W, \H \right).
\end{equation}
\edt{Note that this expectation is the same quantity as $\Mm_n\syn\coefh_n$ used in Eq.\,\eqref{eq:em2.from.em1}.
Using the expression for $\coefh_n$ from Eq.\,\eqref{eq:e-step.1:mean} yields:}
\begin{equation}\label{eq:am.signal.update}
	\fmissh_n
	= \Mm_n\syn\D_n\adjoint{\syn}\transp{\Mr}_n
	\inv{\left(\Mr_n\syn\D_n\adjoint{\syn}\transp{\Mr}_n\right)}
	\fobs_n.
\end{equation}
The whole signal frame, including both the estimated missing samples $\fmissh_n$ and the observed samples $\fobs_n$, can be merged together as:
\begin{equation}\label{eq:am.signal.update.whole}
	\fsigh_n = \syn\D_n\adjoint{\syn}\transp{\Mr}_n
	\inv{\left(\Mr_n\syn\D_n\adjoint{\syn}\transp{\Mr}_n\right)}
	\fobs_n.
\end{equation} 

\subsection{NMF parameters update}

For the NMF model update, we aim at deriving the computation directly from the optimization problem \eqref{eq:mle.2}.
Since not only the observed samples but also the (estimated) missing ones are fixed in this step,
this is equivalent to minimizing
\begin{equation}
	-\log p\left(
	\fsig_n
	\mid\W,\H
	\right)
	= \log{\det{\pi\syn\D_n\adjoint{\syn}}}
	+ \transp{\fsig_n}
	\inv{\left(\syn\D_n\adjoint{\syn}\right)}
	\fsig_n
\end{equation}
with respect to $\W,\H$. To simplify the development of the method, let us pose the following assumption.

\begin{assumption}[invertibility of the synthesis]\label{ass:invertible}
	The synthesis operator $\syn$ is invertible and the analysis operator is $\ana = \inv{\syn}$.
	In particular, this means that $\syn$ is square, i.e., $\Fdim = \Wdim$.
\end{assumption}

Under assumption \ref{ass:invertible}, we see that $\det{\pi\syn\D_n\adjoint{\syn}} = \pi^\Wdim\det{\syn}^2\det{\D_n}$ and ${\inv{(\syn\D_n\adjoint{\syn})} = \adjoint{(\inv{\syn})}\inv{\D_n}\inv{\syn}}$.
The optimization problem then reduces to:
\begin{equation}\label{eq:am.invertible}
	\argmin_{\W,\H} \; \log\det{\D_n} + \adjoint{\left(\inv{\syn}\fsig_n\right)}\inv{\D_n}\left(\inv{\syn}\fsig_n\right).
\end{equation}
Now recall that $\D_n = \diag \left([v_{fn}]_{f=1,\dots,\Fdim} \right)$, thus we can rewrite the objective function as:
\begin{equation}\label{eq:is-nmf}
	\log\prod_{f=1}^{\Fdim} v_{fn} + \sum_{f=1}^{F} \left(\inv{\syn}\fsig_n\right)_f^*\frac{1}{v_{fn}}\left(\inv{\syn}\fsig_n\right)_f
	=
	\sum_{f=1}^{\Fdim} \log v_{fn} + \sum_{f=1}^{\Fdim}\frac{\abs{\left(\inv{\syn}\fsig_n\right)_f}^2}{v_{fn}}.
\end{equation}
It is now straightforward to show that the minimization of \eqref{eq:is-nmf} is equivalent to the minimization of the Itakura--Saito divergence:
\begin{equation}
	\argmin_{v_{fn}} \sum_{f = 1}^{F} d_\textup{IS}\left(\abs{\left(\inv{\syn}\fsig_n\right)_f}^2\mid v_{fn}\right),\quad v_{fn} = \sum_{k} w_{fk}h_{kn}.
\end{equation}
Taking into account all the frames finally leads to the desired result that $\W,\H$ are obtained by minimizing $D_\mathrm{IS}(\P\mid\W\H)$ where $p_{fn} = \abs{(\inv{\syn}\fsigh_n)_f}^2$ and $\fsigh_n$ is the signal estimate from Eq.\,\eqref{eq:am.signal.update.whole}.
The procedure is summarized in Alg.\,\ref{alg:am}.

\begin{algorithm}[t]
    \label{alg:am}
	\small
	\DontPrintSemicolon
	\KwIn{reliable samples $\{\fobs_n\}_{n=1,\dots,N}$, respective selection matrices $\{\Mr_n\}_{n=1,\dots,N}$, invertible linear transform $\syn\in\CC^{\Wdim\times\Fdim}$}
	\vspace{0.5em}
	initialize $\W\in\RR^{\Fdim\times K}$, $\H\in\RR^{K\times N}$ non-negative\\
	\Repeat{convergence criteria met}
	{
		\setstretch{1.5}
		\tcp{Signal update:}
		\For{$n = 1,\dots,N$}
		{
			$\D_n \leftarrow \diag \left([v_{fn}]_{f=1,\dots,\Fdim} \right)$ with
			$[v_{fn}]_{f=1,\dots,\Fdim}$ being the $n$-th column of the matrix $\matr{V} = \W\H$
			\\
			$\coefh_n \leftarrow \D_n \adjoint{\syn}\transp{\Mr}_n
			\inv{\left(\Mr_n\syn \D_n \adjoint{\syn}\transp{\Mr}_n\right)}
			\fobs_n$ \label{alg:coef}
			\\
			$\fsigh_n \leftarrow \inv{\syn}\coefh_n$
			\\
			$p_{fn} \leftarrow \abs{(\coefh_n)_f}^2,\; f = 1,\dots,\Fdim$
		}
		\tcp{Model update:}
		\Repeat{satisfied with the factorization}
		{
			$\W \leftarrow \W \odot \cfrac{%
				\big((\W\H)^{\odot[-2]} \odot \P\big) \transp{\H}
			}{%
				(\W\H)^{\odot[-1]} \transp{\H}
			}$
			with $\P = [p_{fn}]\hspace{-10em}$
			\\[0.5em]
			$\H \leftarrow \H \odot \cfrac{%
				\transp{\W}\big((\W\H)^{\odot[-2]} \odot \P\big)
			}{%
				\transp{\W}(\W\H)^{\odot[-1]}
			}$
			with $\P = [p_{fn}]\hspace{-10em}$
			\\[0.5em]
			normalize columns of $\W$,
			scale rows of $\H$\\[0.5em]
		}
	}
	\vspace{0.5em}
	\KwOut{$\FSIGH =  \left[ \fsigh_1,\dots,\fsigh_N \right], \hat\W = \W, \hat\H = \H$}
	\caption{Audio inpainting via \AM.}
\end{algorithm}

\begin{remark}
	A simple heuristic possibility for the case of non-invertible $\syn$ is to compute the spectrum of $\fsigh_n$, defined in Eq.\,\eqref{eq:am.signal.update.whole}, as
	\begin{equation}
		\coefh_n = \ana\fsigh_n = \ana\syn\D_n\adjoint{\syn}\transp{\Mr}_n
		\inv{\left(\Mr_n\syn\D_n\adjoint{\syn}\transp{\Mr}_n\right)}
		\fobs_n
	\end{equation}
	and the power spectrogram $p_{fn} = \abs{(\coefh_n)_f}^2$.
	Then, we apply the multiplicative rules to minimize $D_\mathrm{IS}(\P\mid\W\H)$.
	However, this approach is not justified by the minimization of \eqref{eq:is-nmf} with respect to $\W,\H$.
	The problem is that if we cannot compute the inversion $\inv{(\syn\D_n\adjoint{\syn})}$ as $\matr{A}\inv{\D_n}\matr{B}$ for some matrices $\matr{A},\matr{B}$, we cannot separate the individual diagonal entries of $\D_n$ to fit it to the IS-NMF problem.
\end{remark}

\begin{remark}
	Note that in the setting imposed in Assumption \ref{ass:invertible} (\nameref{ass:invertible}), \EMTF{} is equivalent to \EMT, but the alternating minimization produces a different algorithm, because we do not include any covariance matrix in the power spectra.
	Thus, \AM{} might provide a different inpainting solution, as demonstrated by the numerical experiments in Section \ref{sec:experiments}.
\end{remark}

\begin{remark}[computational complexity]\label{rem:computational.complexity}
    It is intricate to express the computational complexity of the algorithms \EMTF, \EMT{} and \AM.
    However, several claims are evident:
    \begin{enumerate}
        \item In \EMTF{} on line \ref{alg:sigma} of Alg.\,\ref{alg:emtf}, we do not in fact need to compute the whole matrix $\bSigmah_n$, but solely its diagonal entries.
        Thus, if the complicated term $\D_n\adjoint{\syn}\transp{\Mr}_n\inv{\left(\Mr_n\syn \D_n \adjoint{\syn}\transp{\Mr}_n\right)}$ is computed and saved on line \ref{alg:coef} (e.g., using Matlab's \href{https://www.mathworks.com/help/matlab/ref/mrdivide.html}{mrdivide} without the need to perform the matrix inversion), then no more matrix multiplications are needed on line \ref{alg:sigma} since the diagonal entries can be extracted by computing only $F$ scalar products.
        \item In \AM{}, the matrix $\D_n\adjoint{\syn}\transp{\Mr}_n\inv{\left(\Mr_n\syn \D_n \adjoint{\syn}\transp{\Mr}_n\right)}$ does not need to be calculated at all -- we can first compute $\inv{\left(\Mr_n\syn \D_n \adjoint{\syn}\transp{\Mr}_n\right)}\fobs_n$ efficiently
        and then multiply the resulting vector with the matrix $\D_n\adjoint{\syn}\transp{\Mr}_n$.
        The difference to \EMTF{} is that the inversion operates with a vector, not a matrix, which makes \AM{} less computationally demanding per iteration than \EMTF{}.
        \item Although a similar strategy as in \EMTF{} can be applied to \EMT, computing even the diagonal entries of $\bSigmah_n^{\textup{alt}}$ as derived in Eq.\,\eqref{eq:e-step.2} is more complicated due to the multiplication with $\ana\syn$.
        As a result, \EMT{} has computationally the most expensive iteration from the three algorithms, even in settings when the operations can be implemented efficiently using FFT.
    \end{enumerate}
\end{remark}

\section{Experiments}
\label{sec:experiments}

In this section, we evaluate the performance of the proposed estimators for the task of inpainting noise-less musical recordings. The implementation
\edtlang{was} 
done in Matlab with the use of the LTFAT toolbox \cite{LTFAT}. For the sake of reproducibility, the source code is published online.\footnote{\url{https://github.com/ondrejmokry/InpaintingNMF}}


\edt{We first present}
\edt{an experiment testing the restoration of missing individual samples} 
at random location \edt{(see e.g.\ \cite{Lieb2018:Audio.Inpainting} for similar use case)}.
\edt{This setup served as a demonstrative example
for the comparison of the estimators on a single signal only,}
which
\edtlang{was} 
mainly motivated by the high computational cost of the algorithms (see Section \ref{sssec:random.missing.samples}).
\edt{Then we considered a more challenging setting of compact short to medium gaps.}
\edt{We evaluated the estimators against each other in the gap-filling experiment (Section \ref{sssec:missing.gaps}),
and finally we compared them to state-of-the-art baselines (Section \ref{ssec:comparison.soa}).}


\subsection{\edt{Evaluation metrics}}
\label{ssec:evaluation.metrics}

As a measure of audio quality, we
\edtlang{considered} 
the 
signal-to-noise ratio (SNR) defined as \cite{Adler2012:Audio.inpainting}
\begin{equation}
    \textup{SNR}(\sig,\sigh) = 10\log_{10}\frac{\norm{\sig}^2}{\norm{\sig-\sigh}^2},
\end{equation}
and expressed in dB, as well as the perceptually motivated objective measure PEMO-Q \cite{Huber:2006a}.
All the metrics measure the similarity
\edt{between} 
the ground-truth signal $\sig$ and its estimate $\sigh$; SNR represents the sample-wise similarity
\edt{between} 
the waveforms, whereas PEMO-Q 
estimates the perceived difference (objective difference grade, ODG) on a scale ranging from $-4$ (very annoying) to $0$ (imperceptible).
Note that since we
\edtlang{assumed} 
a noise-less scenario and all the methods considered fit the reliable samples perfectly, we measure the SNR only on the inpainted segments.

The relative solution change, used as a measure of convergence, is computed as ${\norm{\sigh^{(i+1)}-\sigh^{(i)}}/\norm{\sigh^{(i)}}}$, where $\sigh^{(i)}$ is the estimate at the $i$-th iteration folded together from $\syn\COEFH = \syn[\coefh_1,\dots,\coefh_N] = [\fsigh_1,\dots,\fsigh_N]$, where $\COEFH$ is the output of the proposed algorithms.
Similarly, we
\edtlang{computed} 
the relative objective change, where the objective is the negative log-likelihood of Eq.\,\eqref{eq:mle.1} for \EMTF{} and \EMT{} and of Eq.\,\eqref{eq:mle.2} for \AM, evaluated using the current iterates of $\W$, $\H$ and the corresponding solution.

\subsection{Comparison of the proposed estimators}
\label{ssec:comparison.of.the.estimators}

We
\edtlang{started} 
the experiments with the comparison of the behavior of the three estimators, \EMTF, \EMT{} and \AM.
In \ref{sssec:random.missing.samples}, we
\edtlang{present} 
a preliminary comparison including the very demanding calculation of the objective function.
Then, we
\edtlang{give a validation of} 
the results in \ref{sssec:missing.gaps} for the gap-filling task.

\subsubsection{\edt{Introductory} comparison with random missing samples}
\label{sssec:random.missing.samples}

First, we illustrate the capabilities of the three estimators \EMTF, \EMT{} and \AM{} on a demonstrative example of an inpainting problem.
The original signal is an excerpt of the first 12 seconds of the song \emph{Mamavatu} by Susheela Raman, containing acoustic guitar and drums, sampled at 16\,kHz.
We
\edtlang{discarded} 
60\,\% of the signal samples (chosen randomly) and
\edtlang{performed} 
inpainting using \EMTF, \EMT{} and \AM.
The temporal frames
\edtlang{were} 
extracted using sine window of length $\Wdim = 1024$ samples (64\,ms), the hop length
\edtlang{was set to} 
512 samples and we
\edtlang{used} 
$K = 20$ components of the NMF.
To distinguish between the individual algorithms,
\edt{we used DFT as the transform $\adjoint{\syn}$ with} the number of frequency channels $\Fdim$
\edt{varying}
between $\Wdim$ and $2\Wdim$, which corresponds to examples \ref{itm:T.is.unitary} and \ref{itm:T.is.synthesis} of subsection \ref{ssec:equivalence}, respectively.
\edt{The inner loop for NMF was run for 10 iterations.}
\edt{%
\begin{remark}%
    As already mentioned in Remark \ref{rem:dft.independence}, choosing the DFT and its inverse transform as $\syn$ and $\adjoint{\syn}$, respectively, causes violation of the independence assumption \ref{ass:gaussian}.
    A natural way of overcoming this issue would be to work with half of the frequency spectrum and only compute the complex-conjugate half when synthesizing the resulting signal.
    However, this approach is not practical since all our algorithms require to compute the inverse $\inv{\left(\Mr_n\syn \D_n \adjoint{\syn}\transp{\Mr}_n\right)}$, which is quite involved.
    Therefore, we used the whole frequency spectrum throughout the experiments.
    To ensure that the resulting signal are real-valued, we initialized $\W$ such that its columns exhibit the desired symmetry. 
    It is straightforward to show that the symmetry is preserved throughout all the operations during the updates of our algorithms (element-wise addition and multiplication of such symmetrical matrices, right multiplication with a real-valued matrix, etc. - see  \cite[Sec.\,III.D]{VialMagronOberlinFevotte2021:Phase.retrieval.with.Bregman.divergences} for more details).
\end{remark}
}

\begin{figure}
    \centering
    \scalebox{0.51}{\input{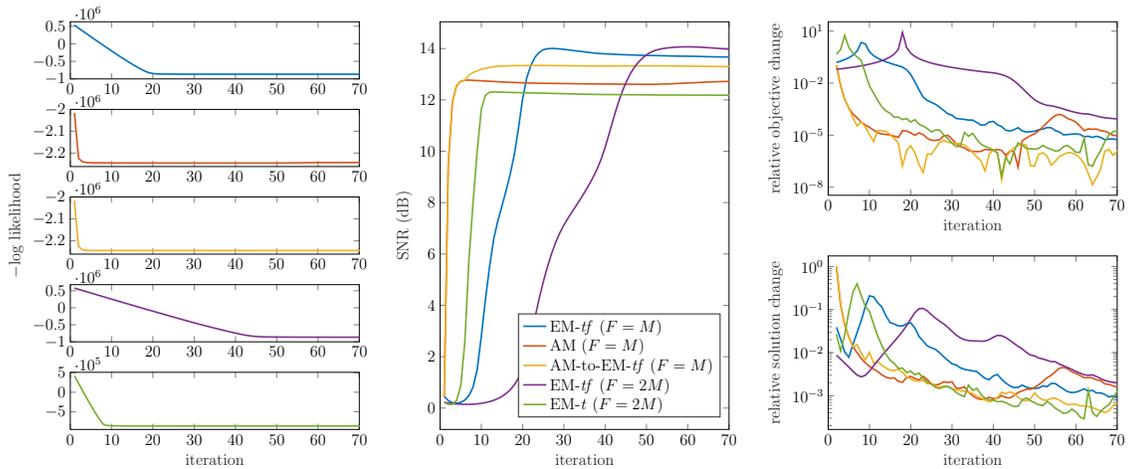}}
    \caption{%
    Comparison of the performance and convergence properties of \EMTF, \EMT{} and \AM, including the switching variant \AMEM.
    The legend in the middle plot is common for the whole figure.
    Note that the first column thus shows three different quantities, since the objective depends on the choice of $\Fdim$ and also on the algorithm.
    Especially, the formula for log likelihood switches after the initializing iterations of \AMEM, which is however disregard on purpose in the plot.}
    \label{fig:emtf_emt_am}
\end{figure}

The comparison is visualized in Fig.\,\ref{fig:emtf_emt_am} by means of several quantities: the negative log-likelihood (i.e., the objective function of the estimation problem \eqref{eq:mle.1} or \eqref{eq:mle.2}), SNR, relative objective change and relative solution change,
\edt{as introduced in Subsection \ref{ssec:evaluation.metrics}.}
Our key finding is the difference in the observed convergence speed with respect to iteration count.
It is visible in all the plots of Fig.\,\ref{fig:emtf_emt_am} that \AM{}
\edtlang{approached} 
its solution
faster than \EMTF{}, which is even more pronounced with respect to CPU time according to remark \ref{rem:computational.complexity} (\nameref{rem:computational.complexity}).
However, the quality of the solution \redt{was} \edt{slightly worse and it} 
\redt{decreased after further iterations}.
A similar behavior
\edtlang{was} 
observed for the case of redundant transform $\syn$.
This redundancy causes that:
\begin{enumerate}
    \item the convergence of \EMTF{} for $\Fdim = 2\Wdim$
    \edtlang{was} 
    slower than with $\Fdim = \Wdim$ while reaching similar reconstruction quality,
    \item the convergence of \EMT{} for $\Fdim = 2\Wdim$
    \edtlang{was} 
    faster than both cases of \EMTF, but as in the case of \AM, it
    \edtlang{reached} 
    worse reconstruction quality.
\end{enumerate}

Based on these first observations, a natural question arises: Can we combine the convergence properties of \AM{} with the performance of \EMTF?
To answer it, we consider a combined algorithm \AMEM{}, which consists in initializing \EMTF{} with $5$ iterations of \AM{}.
As observed in Fig.\,\ref{fig:emtf_emt_am}, the initialization with \AM{}
\edtlang{improved} 
the algorithmic behavior
in the first few iterations and at the same time, the resulting restoration quality
\edtlang{was} 
the same as with \EMTF.
However, the number of iterations needed to reach the peak performance
\edtlang{was} 
not reduced by the switching strategy.

\subsubsection{Missing gaps}
\label{sssec:missing.gaps}

To validate the
\edt{previous} 
results concerning \EMTF{} and \AM, we
\edtlang{performed} 
a larger experiment with a set of signals and for the more demanding problem of inpainting short to middle-length gaps (instead of random subsampling).
%
\edt{
We
\edtlang{used} 
the set of 10 musical recordings from the EBU SQAM dataset \cite{EBUSQAM,EBU_SQAM_Manual}, sampled at 44.1\,kHz and shortened to 7 seconds, as used commonly in recent related publications~\cite{MokryRajmic2020:Inpainting.revisited,TaubockRajbamshiBalasz2021:SPAINMOD}.
}

In each of the 10 signals from the EBU SQAM dataset, 10 gaps of given length
\edtlang{were} 
artificially introduced, with the gap length ranging from 20 to 80\,ms.
We
\edtlang{did} 
not track the objective function in this case, because its computation 
\redt{was} computationally \edtlang{too} demanding. We also
\edtlang{focused} 
on the practical case of $\Fdim = \Wdim$ and invertible transform $\syn$ representing the \edt{inverse} DFT in each temporal frame, thus \EMT{}
\edtlang{was} 
omitted.
The frame length
\edtlang{was set to} 
$\Wdim = 4096$ samples (approx.\ 92\,ms), and the temporal frames
\edtlang{were} 
extracted using sine window with 50\% overlap.
\edt{We used $K=20$ components of the low-rank model according to assumption \ref{ass:nmf},
and the algorithms \ref{alg:emtf} and \ref{alg:am} were run with maximum number of iterations set to 100,
while the inner cycle for the NMF used 10 iterations.}

\begin{figure}
    \centering
    \scalebox{0.55}{\input{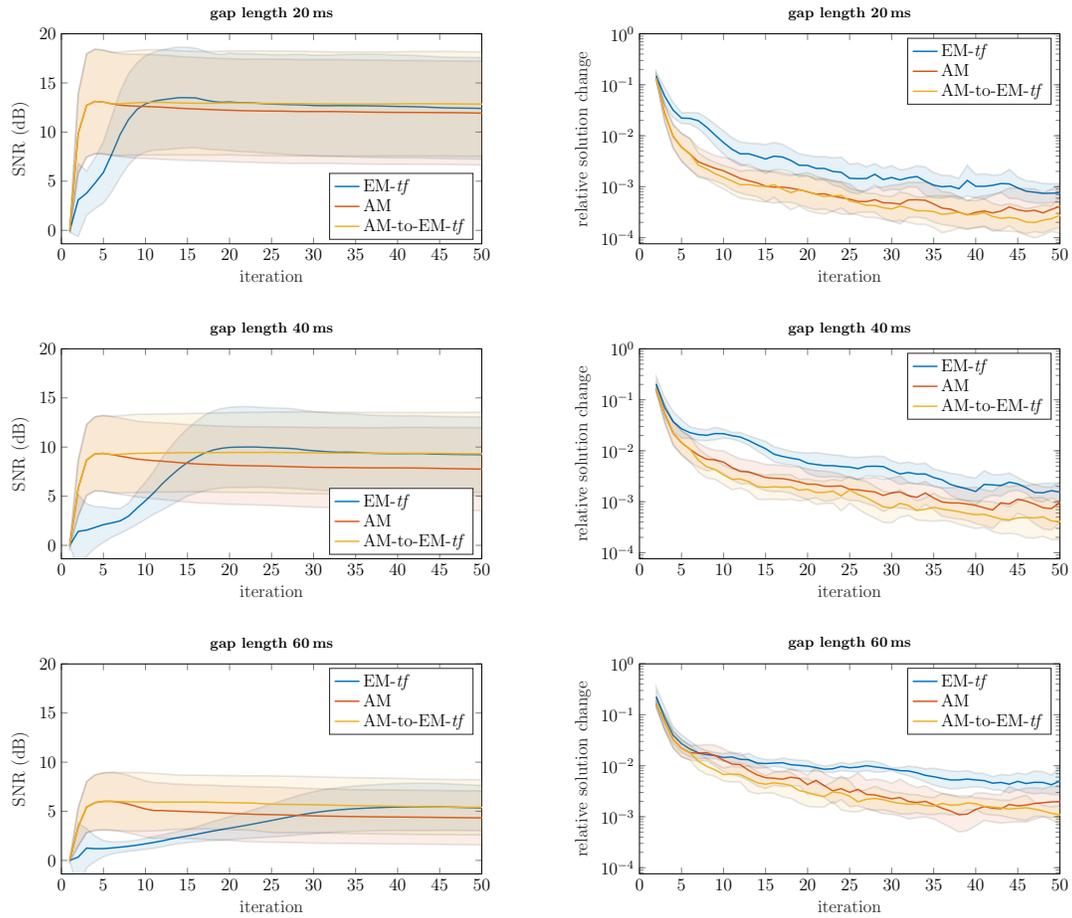}}
    \caption{%
    Comparison of the performance of \EMTF{}, \AM{}, and the switching variant \AMEM{} (switching after 5 iterations).
    The left column shows the evolution of the SNR over iterations, the right column shows the relative solution change.
    Both the metrics are averaged over the dataset and plotted together with 95\% confidence interval represented by the light colored areas (note that for the relative solution change plot, this confidence interval is computed from the decimal logarithm of the data).
    }
    \label{fig:em_am}
\end{figure}

As in the previous experiment, we
\edtlang{observed} 
that the performance difference between \AM{} and \EMTF{}
\edtlang{was} 
not significant, but \AM{}
\edtlang{reached} 
its peak faster, as shown in the plots for SNR in Fig.\,\ref{fig:em_am}.
The relative solution change supports this observation, as it has been demonstrated above (see Fig.\,\ref{fig:emtf_emt_am}) that this measure mostly corresponds to the convergence of the algorithm with respect to its objective value.
A new observation is that this phenomenon depends on the gap length -- the longer the gap, the slower the convergence of \EMTF{} is, compared to \AM.

\begin{figure}[ht]
    \centering
    \scalebox{0.6}{\input{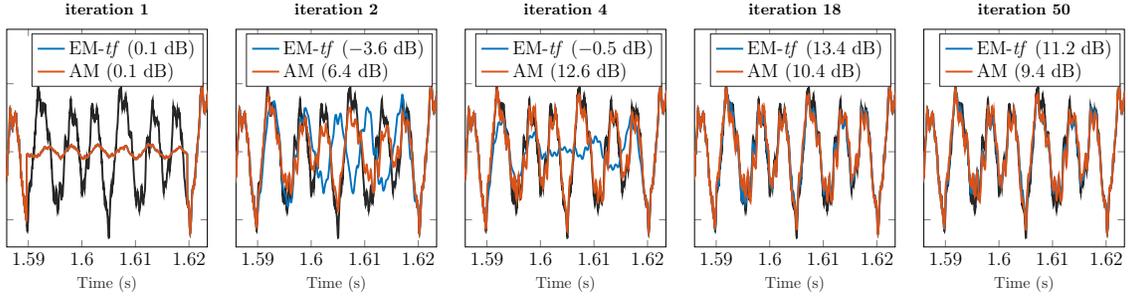}}
    \caption{\edt{Example of the difference between \EMTF{} and \AM{} during iterations, namely for the guitar recording from our testing set and gap length 30\,ms.
    The waveforms are compared to the undegraded original waveform (in black on the plots).
    The legend indicates SNR values \emph{for the whole signal} and for each method at the corresponding iteration number.
    The iterations are chosen such that the peak solution (in terms of SNR) is displayed both for \EMTF{} and \AM{} (iterations 18 and 4, respectively).
    Also note that the solution in the first plot is the same for \EMTF{} and \AM{}, since both algorithms use the same initialization, thus there is no difference in their estimate after 1 iteration (\textit{cf.} line 5 of Alg.\,\ref{alg:emtf} and \ref{alg:am}).
    }
    }
    \label{fig:single_gap}
\end{figure}

\edt{To provide an example supporting the observation about convergence properties of \EMTF{} and \AM{}, we present the evolution in the temporal domain in Fig.\,\ref{fig:single_gap}.
We have taken a single signal from the gap-filling experiment and studied how the temporal solution changes during iterations inside a particular gap.
The observation is consistent with 
\redt{that obtained from} Fig.\,\ref{fig:em_am}.
We see that \AM{} already identified the basic shape of the original signal in iteration 2 and then only improved the estimate in a few iterations.
On the other hand, the first estimate of \EMTF{} in iteration 2 seems to be almost the opposite, compared to the original signal.
As a result, \EMTF{} first returned to the solution being around zero, before finally reaching a quality estimate in later iterations.}

\edt{The interpretation here is that the inclusion of the covariance matrices $\bSigmah_n$ in \EMTF{} helps to fit the model closer to the original signal, but only when the current estimate of the model (matrices $\W,\H$) is already relevant.
This corresponds to the observation that the initial estimates of the signal via \EMTF{} were observed to be inferior compared to \AM{}.
It is also consistent with Fig.\,\ref{fig:em_am}, which shows that the decrease observed in the performance of \AM{} could be prevented by including the covariance matrices into the estimation (i.e.\ by switching to \EMTF{}) if the estimate is already sufficient to some extent.}

\subsection{Comparison with the state of the art}
\label{ssec:comparison.soa}

Finally, we
\edtlang{compared} 
our NMF-based methods to the following state-of-the-art techniques for the task of inpainting middle-length audio gaps,
\edt{based on recent publications \cite{MokryRajmic2020:Inpainting.revisited,TaubockRajbamshiBalasz2021:SPAINMOD,Rajbamshi2021:Audio.inapinting.ell.1.dictionary.learning}}:

\begin{itemize}
    \item Janssen's AR approach \cite{javevr86} (denoted \textbf{Janssen}): This iterative algorithm builds upon a frame-wise AR nature of the clean signal.
    At each iteration, it estimates the AR coefficients of the signal estimate (starting from the observed signal with the missing samples initialized as zeros), and then recompute the missing samples using the reliable samples and the current model parameters' estimates.
    \item Modified variant of Janssen's algorithm (\textbf{Janssenmod}): This method is similar to the AR approach, but instead of processing the signal in overlapping segments, it treats each gap with its context individually.
    \item Cross-faded extrapolation based on AR modeling \cite{Etter1996:Interpolation_AR} (\textbf{LR}): This is an efficient method for inpainting of compact gaps with sufficiently long reliable contexts.
    It consists of estimating two AR models for the left and right contexts, extrapolating the contexts into the gap using the fitted model and cross-fading the two candidate solutions.
    \item \textbf{SPAINlearned} \cite{TaubockRajbamshiBalasz2021:SPAINMOD}, a variant of the sparsity-based non-convex approach SPAIN \cite{MokryZaviskaRajmicVesely2019:SPAIN}, which treats individual gaps and their contexts (instead of overlapping frames as in SPAIN).
    The performance is further enhanced by a dictionary learning step to deform the STFT based on the particular signal such that it allows for a sparser representation than using the STFT.
    \item A convex alternative is the (weighted) $\ell_1$ minimization as a relaxation of the non-convex sparsity \cite{MokryRajmic2020:Inpainting.revisited} (\textbf{reweighted}).
\end{itemize}

For \redt{the} 
methods which use STFT or segmentation (\textbf{Janssen}, \textbf{SPAINlearned}, \textbf{reweighted}, and the NMF-based methods),
\edtlang{we used} 
a sine window of length 4096 samples (approx.\ 92\,ms) with 50\% overlap.
The NMF-based methods
\edtlang{were} 
applied with $\Wdim = \Fdim$, thus \EMT{}
\edtlang{was} 
equivalent to \EMTF{} and
\edtlang{was therefore} 
omitted for brevity.
The AR-based methods
\edtlang{used} 
a model of order 512.
The context of \textbf{Janssenmod} and \textbf{LR}
\edtlang{was} 
set to 4096 samples, while \textbf{SPAINlearned}
\edtlang{used} 
a longer context (8192 samples) for the sake of the dictionary learning. These values
\edtlang{were} 
chosen based on the corresponding studies, where they have shown good performance.
For particular choices of all the parameters of the individual methods, please refer to the published source code.

\begin{figure}[ht]
    \centering
    \scalebox{0.6}{
%
%
%
\definecolor{mycolor1}{rgb}{0.00000,0.44700,0.74100}%
\definecolor{mycolor2}{rgb}{0.85000,0.32500,0.09800}%
\definecolor{mycolor3}{rgb}{0.92900,0.69400,0.12500}%
\definecolor{mycolor4}{rgb}{0.49400,0.18400,0.55600}%
\definecolor{mycolor5}{rgb}{0.46600,0.67400,0.18800}%
\definecolor{mycolor6}{rgb}{0.30100,0.74500,0.93300}%
\definecolor{mycolor7}{rgb}{0.63500,0.07800,0.18400}%
\def\NMFlinethickness{0.85mm}
\begin{tikzpicture}

\begin{axis}[%
width=3.75in,
height=3.25in,
at={(0.0in,0.0in)},
scale only axis,
xmin=20,
xmax=80,
xtick={20,30,...,80},
xlabel style={text=white!15!black},
xlabel={gap length (ms)},
ymin=0,
ymax=14,
ylabel style={text=white!15!black},
ylabel={SNR (dB)},
axis background/.style={fill=white},
legend style={legend cell align=left, align=left, draw=white!15!black}
]
\addplot [color=mycolor1,line width=\NMFlinethickness]
  table[row sep=crcr]{%
20	13.782953297377\\
30	11.8965512995374\\
40	10.2267773718598\\
50	8.37877274642174\\
60	5.78001337324034\\
70	4.1070733938746\\
80	3.338605230901\\
};
\addlegendentry{\EMTF{}}

\addplot [color=mycolor2,line width=\NMFlinethickness]
  table[row sep=crcr]{%
20	13.4219122857239\\
30	11.3238135303734\\
40	9.63409686772784\\
50	7.4764493492823\\
60	6.43801328979986\\
70	4.69827436692887\\
80	3.84634145222867\\
};
\addlegendentry{\AM{}}

\addplot [color=mycolor3]
  table[row sep=crcr]{%
20	11.5408224859766\\
30	10.1804454021908\\
40	8.96665464251497\\
50	6.76546730997628\\
60	5.38802637532399\\
70	3.42882594265129\\
80	2.85325559752452\\
};
\addlegendentry{Janssen}

\addplot [color=mycolor4]
  table[row sep=crcr]{%
20	11.5651838347964\\
30	10.8526683905199\\
40	9.82614955149836\\
50	8.73441059629002\\
60	8.21776820286809\\
70	7.76807061761876\\
80	6.45098715236429\\
};
\addlegendentry{Janssenmod}

\addplot [color=mycolor5]
  table[row sep=crcr]{%
20	8.68432623554607\\
30	7.95906252106424\\
40	6.86361553161148\\
50	6.06662261617966\\
60	5.47337591702458\\
70	4.95773571732389\\
80	4.30617131237165\\
};
\addlegendentry{LR}

\addplot [color=mycolor6]
  table[row sep=crcr]{%
20	9.66429993898978\\
30	9.20151248405106\\
40	7.92438077855388\\
50	7.51457757007766\\
60	7.00240137646707\\
70	6.18322204391512\\
80	5.54315565543537\\
};
\addlegendentry{SPAINlearned}

\addplot [color=mycolor7]
  table[row sep=crcr]{%
20	10.1466069940243\\
30	7.87367657503244\\
40	6.0564287611885\\
50	5.14088512802542\\
60	4.23694108408959\\
70	3.60253945114359\\
80	2.58389706610786\\
};
\addlegendentry{reweighted}

\end{axis}

\begin{axis}[%
width=3.75in,
height=3.25in,
at={(4.75in,0.0in)},
scale only axis,
xmin=20,
xmax=80,
xtick={20,30,...,80},
xlabel style={text=white!15!black},
xlabel={gap length (ms)},
ymin=-3.79808425486142,
ymax=-0.5,
ylabel style={text=white!15!black},
ylabel={PEMO-Q ODG},
axis background/.style={fill=white},
legend style={legend cell align=left, align=left, draw=white!15!black}
]
\addplot [color=mycolor1,line width=\NMFlinethickness]
  table[row sep=crcr]{%
20	-1.0419914399671\\
30	-1.6551333615367\\
40	-2.09043479649572\\
50	-2.64259626599174\\
60	-2.99332959386921\\
70	-3.36366518733452\\
80	-3.56430272063155\\
};

\addplot [color=mycolor2,line width=\NMFlinethickness]
  table[row sep=crcr]{%
20	-0.994890180276015\\
30	-1.66392201065946\\
40	-2.06223308907955\\
50	-2.63540064412039\\
60	-3.04479798774042\\
70	-3.30537403712524\\
80	-3.46437114456889\\
};

\addplot [color=mycolor3]
  table[row sep=crcr]{%
20	-0.771406941840412\\
30	-1.25417719698379\\
40	-1.73779022602664\\
50	-2.17807546314302\\
60	-2.78765150604516\\
70	-3.25788658004592\\
80	-3.43449403651178\\
};

\addplot [color=mycolor4]
  table[row sep=crcr]{%
20	-1.29549886774632\\
30	-1.87620370701875\\
40	-2.20568091030222\\
50	-2.55097839739479\\
60	-2.78403447348591\\
70	-2.88909052859767\\
80	-3.05388803315659\\
};

\addplot [color=mycolor5]
  table[row sep=crcr]{%
20	-1.84733671252234\\
30	-2.56278182895831\\
40	-2.92061737532089\\
50	-3.13614132698354\\
60	-3.24239278718813\\
70	-3.36179732485814\\
80	-3.43845666323798\\
};

\addplot [color=mycolor6]
  table[row sep=crcr]{%
20	-1.33678817044282\\
30	-1.89565519494565\\
40	-2.19730604471932\\
50	-2.49945294763209\\
60	-2.79461839962118\\
70	-3.08740614079011\\
80	-3.23934011915879\\
};

\addplot [color=mycolor7]
  table[row sep=crcr]{%
20	-1.36672335627334\\
30	-2.21151245174463\\
40	-2.79890852620534\\
50	-3.10468882787204\\
60	-3.28970146085707\\
70	-3.4470587602059\\
80	-3.60007732784135\\
};

\end{axis}

\end{tikzpicture}%
}
    \caption{Comparison with the state-of-the-art algorithms for inpainting short to middle-length gaps.
    The legend is common for both plots.}
    \label{fig:comparison}
\end{figure}
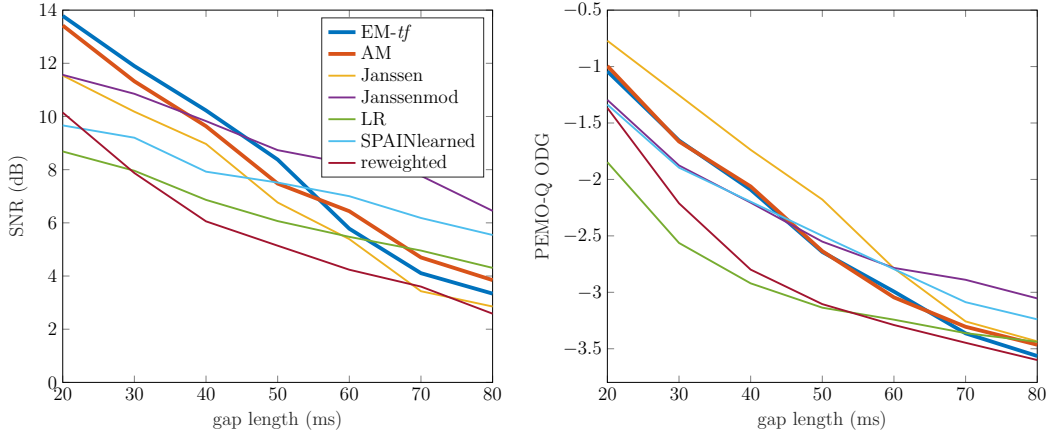

The performance results averaged over the 10 test signals are shown in Fig.\,\ref{fig:comparison}.
We observe that in terms of SNR the proposed NMF-based methods
\edtlang{outperformed} 
the state-of-the-art for short gaps (up to \redt{40}\,ms) while their performance
\edtlang{dropped} 
for longer gaps.
\edt{Note that Janssen 
\redt{exhibited} a~similar behavior, suggesting that the performance 
\redt{was} undermined by estimating the missing samples via Wiener filtering (as mentioned in Sec.\,\ref{sec:formulation}) from a limited number of samples (which 
\redt{was} more pronounced for long gaps with a large number of missing samples per frame).
Note that similar process 
\redt{appears} within the Janssen algorithm in the signal update step \cite[Sec.\,3.3]{Oudre2018:Janssen.implementation}.
On the other hand, sparsity-based methods tend to loose signal energy within long gaps \cite{MokryRajmic2020:Inpainting.revisited}, which may eventually lead to better SNR results.}

\edt{The perceptually-motivated comparison is realized via PEMO-Q ODG.
Here we}
\edtlang{observed} 
that \EMTF{} and \AM{}
\edtlang{were} 
among the top three methods for short to middle gaps \edt{(up to 40\,ms)}.
\edt{For longer gaps, the performance 
\redt{decreased} similarly to the evaluation based on SNR.}

\begin{figure}[ht]
    \centering
    \scalebox{0.6}{\input{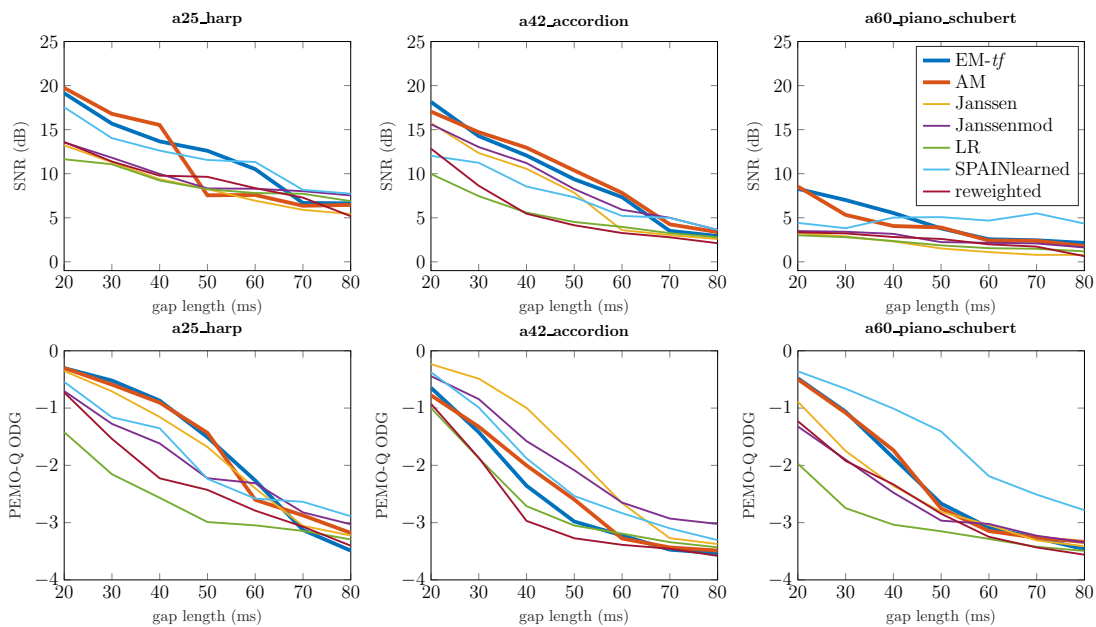}}
    \caption{\edt{Comparison with the state-of-the-art algorithms for inpainting short to middle-length gaps on some selected signals.
    The legend is common for all the plots.}}
    \label{fig:comparison_per_signal}
\end{figure}

\redt{Since the results vary in general, depending on the particular signal,}
\redt{we present a short signal-specific analysis to illustrate which particular results contribute the most to the improvement over the state-of-the-art methods presented in Fig.\,\ref{fig:comparison}.
To this end,}
Fig.\,\ref{fig:comparison_per_signal} focuses on 
\redt{three signals from the dataset on which the NMF-based methods perform particularly well in terms of SNR.}
\redt{While the results in terms of PEMO-Q ODG were mostly consistent with SNR,
we observed a notable difference in the case of the harp signal -- the performance of \AM{} in terms of SNR suddenly dropped for 50\,ms gaps, but the ODG grades stayed on par with \EMTF{}.}

\edt{To summarize the results,
\AM{} reaches a similar performance to \EMTF{} but with fewer iterations (and reduced CPU time given the lower complexity of \AM, see remark \ref{rem:computational.complexity}),
and both methods outperform the state-of-the-art in terms of SNR for small gaps (and rank second in terms of ODG), but they 
\redt{underperformed the other algorithms} for medium to large gaps.}

\section{Conclusion}
\label{sec:conclusion}

In this paper, we derived new estimators for NMF-based audio inpainting.
We formulated inpainting as an optimization problem where the goal is to estimate the signal's power spectrum, which is structured using NMF.
To that end, we have derived three algorithms, among which two are new, which encompasses and extends previous related works~\cite{BilenOzerovPerez2015:declipping.via.NMF,Bilen2018:NTF_audio_inverse_problems}.
Even though the proposed estimators build upon the same low-rank assumption about the signal's TF spectrum, we have shown that there are both theoretical and practical differences between them.
In particular,
they all exhibit a different behavior and lead to different solutions to the audio inpainting problem.
Importantly, the novel approaches (\EMT{} and \AM{}) improve the convergence rate compared to \EMTF{}, while reaching similar reconstruction quality.
\edt{
\redt{Both} NMF-based methods \redt{considered in the comparison} are in general efficient for short to middle gaps (up to 40\,ms), where they 
\redt{provided better results than the state-of-the-art algorithms}.}

Throughout the derivation, we assumed independence of the temporal frames.
A natural extension of the model would be to employ temporal Markov NMF models, as presented e.g., in \cite{Smaragdis2014:Static.dnymic.source.separation.NMF}.
Future research will also study the possibility to leverage psychoacoustics in audio inpainting, e.g., by using perceptually-motivated TF representations \cite{Lieb2018:Audio.Inpainting, Necciari2013:ERBlet}.

\bibliography{literatura.bib}

\end{document}